\documentclass[12pt]{article}
\usepackage[dvips]{graphicx}
\newcommand{\beq}{\begin{equation}}
\newcommand{\eeq}{\end{equation}}
\newcommand{\bdis}{\begin{displaymath}}
\newcommand{\edis}{\end{displaymath}}
\newcommand{\bea}{\begin{eqnarray}}
\newcommand{\eea}{\end{eqnarray}}
\newcommand{\barr}{\begin{array}}
\newcommand{\earr}{\end{array}}
\newcommand{\beas}{\begin{eqnarray*}}
\newcommand{\eeas}{\end{eqnarray*}}

\begin{document}
\begin{center}
{\large{\bf KINEMATICS OF THE THREE  MOVING SPACE CURVES ASSOCIATED WITH
 THE NONLINEAR SCHR\"{O}DINGER EQUATION}}

\vskip .4cm
${\bf Radha~Balakrishnan^1}$~~~{\bf and}~~~
${\bf S.~Murugesh^1}$

\vskip .3cm
\end{center}
\bigskip
\baselineskip18pt
\noindent {\it
 Starting with the  general description of  a moving curve, we have  recently presented a unified 
formalism to show that  three  distinct space curve evolutions can be identified with a given 
integrable equation.  Applying this to the nonlinear Schr\"{o}dinger equation (NLS), we find  three 
sets of  coupled equations for the evolution of the curvature and the torsion, one set for each moving 
curve. The  first set is given by the usual Da Rios-Betchov  equations. The  velocity at each point of 
the curve  that corresponds to this set is  well known to be a  local expression in the curve variables.
In contrast, the velocities of the other two curves are  shown to be nonlocal expressions. Each of the 
three curves is  endowed with a corresponding infinite set of geometric constraints. These  moving 
space curves are found by using their connection with the integrable Landau- Lifshitz equation.  The 
three  evolving curves corresponding to the envelope soliton solution of the NLS  are presented and 
their behaviors compared. }

\vskip 7cm
\hrule
\smallskip
\bigskip
\noindent $^{1}~ {\small\rm The~~ Institute~~ of~~ Mathematical
~~ sciences,~~ Chennai~~ 600~113,~~ India. }$ \\
\newpage
\section{ Introduction}
 The study of possible links between intrinsic kinematics of
 space curves \cite{stru} and integrable soliton bearing equations\cite{ablo}
 deserves attention because of a wide
  variety of applications of moving curves such as
 vortex filament motion in fluids \cite{hasi}, dynamics of continuum
 spin chains \cite{laks},  scroll waves in chemical reactions \cite{keen},
 superfluids \cite{schw}, interface dynamics \cite{gold},
  etc.
 The pioneering work by Hasimoto \cite{hasi} on the motion of
 a vortex filament in a fluid
  was the first to suggest such a link. In classical differential
 geometry, a
 space curve embedded in three dimensions
  is  described
 using the following Frenet-Serret equations \cite{stru}
 for the orthonormal triad of unit vectors made up of its  tangent
 ${\bf t}$, normal ${\bf n}$ and the binormal ${\bf b}$:
\beq\label{eq1}
{\bf t}_s \ = \ K {\bf n} \ ; \ {\bf n}_s  \ =
 \ -K{\bf t} \ + \ \tau
{\bf b}\ ; \ {\bf b}_s \ =\ -\tau {\bf n}.
\eeq
Here,  $s$ denotes the arclength.
The parameters  $K$ and $\tau$ represent the curvature and torsion
 of the space curve. For a moving curve, these will be functions of both $s$ and time $u$.
  The subscript $s$ denotes partial derivative.

 Using the so-called local induction approximation,  Da Rios \cite{dari}
 showed that the curve velocity
 ${\bf v}(s,u)$  at each point $s$ of a vortex filament
(regarded as a moving space curve)
  is given by the  local induction
 equation :
\beq\label{eq3}
{\bf v} = {\bf r}_u=~~K~{\bf b}
\eeq
 For non-stretching curves, using compatibility conditions ${\bf r}_{us}=
{\bf r}_{su}={\bf t}_{u}=K_{s}{\bf b}-K\tau {\bf n}$, along with
 similar compatibility conditions for  ${\bf t}$ and ${\bf n}$ leads to \cite{dari}
\beq\label{eq4}
K_u=-(K\tau)_{s}-K_{s}\tau;~~ \tau_u=[(K_{ss}/K)- \tau^{2}]_{s}-
 KK_{s}.
\eeq
 Subsequently, the above coupled equations for $K$ and $\tau$
 were also derived
 independently by Betchov \cite{betc}, and hence Eq. (\ref{eq4}) is referred to as
 Da Rios-Betchov (DB) equations.
 Connection between  these curve equations and  solitons was
 unravelled by Hasimoto \cite{hasi},
 whose analysis essentially showed that by using the following `Hasimoto function'
\beq\label{eq6}
\psi=~~K~~\exp[i\int~\tau~~ds],
\eeq
  the two equations in (\ref{eq4}) can in fact be combined
  to give the integrable, soliton-bearing nonlinear Schr\"{o}dinger
 equation (NLS)
\beq\label{eq7}
iq_u+q_{ss}+ \frac{1}{2}~|q|^2~q  =0,
\eeq
  with  $q=\psi$. Motivated by this result, Lamb \cite{lamb} gave
 a general procedure
 which helps identify  a certain space curve evolution with a
 given integrable equation. Examples
 such as the NLS, the sine-Gordon equation and the
modified KdV equation were considered. Recently, we presented
 \cite{muru} a unified analysis
 to show that in general, {\it two more} distinct  curve evolutions
 can  also be  so identified.
In this paper,  after outlining this analysis, we specialize to the NLS,
 and  focus on the intrinsic kinematics of
the three moving curves associated with it.
We obtain  coupled evolution equations for the curvature and torsion
 of  each of the two new curves. These are certain analogs of the DB
equations (\ref{eq4}). Each of the three  moving curves  is
 shown to be endowed with an infinite set of
 geometric invariants. Their  natural connection with the
 integrable  Landau-Lifshitz
 equation is pointed out, and  a procedure  to use its solution to
 construct
 the position vectors generating the
 three curves  associated with the NLS is given. As
 an example, the moving curves for an envelope soliton solution
  of the NLS are found, and their behaviors compared.
    Interestingly,  the  velocities  (at every point)  of
 the two new moving  curves  underlying the  general NLS evolution
 turn out to be  certain {\it nonlocal} functions  of the curve
 variables, quite in contrast to the  {\it local} expression (Eq. (\ref{eq3}))
 for the velocity of the  usual moving curve that has  thus far
 been  associated with the NLS.
 Possible application of these results  to vortex filament motion
  in fluids is  suggested.
\section{Identification of three curve evolutions with a given integrable
 equation}
\setcounter{equation}{0}

 To describe  a  moving curve, we find it convenient to introduce
 \cite{radha} the
 following time evolution equations for the Frenet triad $({\bf t}, {\bf n}, {\bf b})$:
\beq\label{eq2}
{\bf{t}}_u\,= g {\bf {n}} +h  {\bf{b}}\ ;\   {\bf{n}}_u \,=
 -  g {\bf{t}} + \tau_{0} {\bf{b}}\ ; \  {\bf{b}}_u\,=\,-h
{\bf{t}} - \tau_{0} {\bf{n}}.
\eeq
The parameters $g,h$ and $\tau_0$ are, at this stage,
general parameters which determine
 the time evolution  of the curve. They are  functions of both $s$
 and $u$. The subscript $u$ stands  for  partial derivative.
 On imposing  the compatibility conditions
\beq
{\bf t}_{su} \ = \ {\bf t}_{us} \ ;
 \ {\bf n}_{su} \ = \ {\bf n}_{us}
\quad ; \quad {\bf b}_{su} \ = \ {\bf b}_{us},
\eeq
and  using Eqs. (\ref{eq1}) and (\ref{eq2}), we obtain
\beq\label{eq8}
K_u=(g_s-\tau h);~~\tau_u=(\tau_0)_s+Kh;~~h_s=(K\tau_0-\tau g).
\eeq
  Lamb's procedure
 \cite{lamb}  will be referred to as
 "formulation (I)", to distinguish it from  two other
   additional formulations that are possible \cite{muru}.
 We shall use the  subscripts $1,2$ and $3$ on the various curve parameters
  appearing in  the three formulations, for ready reference.\\
\noindent {\small {\bf Formulation I:}}
  Here, one  combines the second and third of
  Eqs. (\ref{eq1}), which immediately suggests  the definition  of
  a  complex vector  ${\bf N}=({\bf n}+i{\bf b})~~
 {\rm exp}[i\int \tau~ds]$,
  along with the complex function $\psi$ given in Eq. (\ref{eq6}).
  By writing ${\bf N}_s, {\bf t}_s, {\bf N}_u$ and ${\bf t_u}$
 in terms of ${\bf t}$ and ${\bf N}$, imposing the compatibility
 condition ${\bf N}_{su} \ = \ {\bf N}_{us}$,
and equating the coefficients of {\bf t} and {\bf N} in it,  one obtains
\beq\label{eq9}
\psi_u+\gamma_{1s}+(1/2)[\int~(\gamma_1\psi^*-\gamma_1^*\psi)~~ds]~\psi=0.
\eeq
where
\beq\label{eq10}
\gamma_{1}=-(g+ih)~~{\rm exp}[i\int\tau~ds].
\eeq
Thus for an appropriate choice of $\gamma_1$ as a function of $\psi$
 and its derivatives,
  a known integrable equation for $\psi$
can be obtained from (\ref{eq9}). Then, by comparing
 a known solution of this equation with the Hasimoto function (\ref{eq6}),
  the curvature $K$ and  torsion $\tau$ of the corresponding
 moving space curve can be found. Next, using the above mentioned
 specific  choice of $\gamma_1$  in Eq. (\ref{eq10}),  the curve
 evolution parameters  $g$ and $h$  can be determined as  some specific
 functions of $K$ ,$\tau$ and their derivatives. Armed with these,
 $\tau_0$ can also be found from the third equation in
   (\ref{eq8}). Thus a  set of curve parameters $K$, $\tau$,
 $g$, $h$ and $\tau_0$
 has been found explicitly. In other words, a
   certain moving curve described by these parameters has thus
been identified with  a given solution of the integrable
 equation for $\psi$.

  As already mentioned, we  have recently shown \cite{muru}
 that in addition to the above, there are
 two other distinct curves that can be identified
 with a given  integrable equation. The procedures
  leading to these curves  will be called
 formulations (II) and (III),  respectively.\\
\noindent{\small {\bf Formulation (II):}}
 Combining  the first two
 equations in (\ref{eq1}) appropriately, we see that
 a  complex vector
 ${\bf M}=({\bf n}-i{\bf t})~{\rm exp}[i\int K~ds]$ and a complex function
\beq\label{eq11}
\Phi(s,u)=\tau\exp{[i\int K~ds]},
\eeq
  appear. Proceeding in a fashion analogous to (I) above,
 and imposing  ${\bf M}_{su}={\bf M}_{us}$, we get
\beq\label{eq12}
\Phi_u+\gamma_{2s}+(1/2)[\int~(\gamma_2\Phi^*-\gamma_2^*\Phi)~~ds]~\Phi=0.
\eeq
where
\beq\label{eq13}
\gamma_{2}=-(\tau_{0}-ih)~~{\rm exp}[i\int K~ds].
\eeq

\noindent {\small {\bf Formulation (III):}}
  Here,  we combine  the first and
 third equations of (\ref{eq1}), leading to the appearance of a
   complex vector ${\bf P}=({\bf t}-i{\bf b})$, and
 a complex function $\chi$  given by
\beq\label{eq14}
\chi(s,u)=(K+i\tau).
\eeq
 On  requiring  ${\bf P}_{su}  = {\bf P}_{us}$,  we get
\beq\label{eq15}
\chi_u+\gamma_{3s}+(1/2)[\int~(\gamma_3\chi^*-\gamma_3^*\chi)~~ds]~\chi=0.
\eeq
where
\beq\label{eq16}
\gamma_{3}=-(g+i\tau_{0}).
\eeq
 Now,  Eqs. (\ref{eq12}) and (\ref{eq15}) have the same
 form as Lamb's  equation  (\ref{eq9}).
  The discussion in the paragraph following Eq. (\ref{eq10}) makes it
  clear that for an appropriate choice
 of $\gamma_2$  as a  function of $\Phi$ and its derivatives,
  and of $\gamma_3$  as a  function of $\chi$ and its
 derivatives, known integrable  equations
 for $\Phi$ and $\chi$ can be obtained from these equations.

 It is important to note from Eqs. (\ref{eq6}), (\ref{eq11})
 and (\ref{eq14}) that   the
complex functions $\psi$, $\Phi$ and $\chi$  that satisfy the
  (same) integrable equation
 in the three formulations  are {\it different}
functions of $K$ and $\tau$. Further,  we see that the
 complex quantities  $\gamma_1$, $\gamma_2$ and $\gamma_3$
 (Eqs. (\ref{eq10}),
 (\ref{eq13})
 and (\ref{eq16})) that arise in the
 three formulations respectively,
  also  involve {\it different} combinations  of  the curve
evolution parameters  $g,h$  and  $\tau_{0}$.
 Thus it is clear that
the three formulations yield different sets of
   curve parameters, and hence  describe three  {\it distinct}
 moving space  curves that
 can be associated with a given solution of an integrable
  equation. (We remark that it should be possible to extend
  our analysis to include partially integrable equations as well.)
 In the next section, we  apply these results to the
 NLS, to demonstrate this clearly.\\
\section{The NLS : Two analogs of Da Rios-Betchov equations}
\setcounter{equation}{0}

 From the three formulations discussed in the last section,
 it can be easily verified that the respective choices
\beq\label{eq17}
\gamma_1  = -i \psi_{s};~~\gamma_2= -i \Phi_s;~~\gamma_3= -i \chi_s,
\eeq
when used in Eqs. (\ref{eq9}), (\ref{eq12}) and (\ref{eq15})
lead to the NLS given in (\ref{eq7}), with $q$ identified with
  the three complex functions $\psi, \Phi$ and $\chi$,
 respectively.
  Since the NLS  is a completely integrable
 equation \cite{ablo}, it  possesses
 an infinite set of  conserved quantities
 $I_{k},~ k=1,2....\infty$,  which are in involution pairwise.
 The first three of these invariants are given by \cite{hami}
\beq\label{eq18}
I_{1}=\int~~|q|_{s}^2~~ds;~ I_{2}=~~(1/2i)\int~~(q_{s}q^{*}~-~q_{s}q^{*})~~ds;~
I_{3}=\int~[|q_{s}|^{2}~-~\frac{1}{4}|q|^{2}]~ds;....
\eeq
Let us consider the implications of this in the
 three formulations:\\
{\bf (I)} Setting $q=\psi$ in (\ref{eq7}) and equating real
 and imaginary parts leads to DB equations  (\ref{eq4}).
    Therefore these also possess an infinite
 number of conserved quantities, obtained by setting $q=\psi$ in
 Eq. (\ref{eq18}). The invariants now appear as geometric  constraints \cite{lang}:
\beq\label{eq19}
I_{1}=\int~K^2~~ds;~ I_{2}=\int~K^{2}~\tau~~ds;~
I_{3}=\int~[K_{s}^{2}+K^{2} \tau^{2}-~\frac{1}{4}K^{4}]~~ds;....
\eeq
 Further, an inspection of Eqs. (\ref{eq4})
 shows that  the total {\it torsion} $I_{0}=\int~\tau~ds$ is
also conserved. This additional geometric constraint
 has no counterpart among  the NLS invariants given in (\ref{eq18}).\\
{\bf (II)} Here, we set $q=\Phi$ in (\ref{eq7})
 to yield the coupled equations
\beq\label{eq20}
K_u=[(\tau_{ss}/\tau)- K^{2}]_{s}
-\tau \tau_{s};~~
\tau_{u}=-(K\tau)_{s}-\tau_{s}K.
\eeq
 This is the first analog of the DB equations (\ref{eq4}).
 As is obvious from comparing $\psi$ in (\ref{eq6}) and $\Phi$
  in Eq.(\ref{eq11}), this analog
   may be found by simply interchanging
 $K$ and $\tau$ in  (\ref{eq4}).
   Thus the associated  infinite
 number of  geometric constraints can also be found using this
 interchange in Eq. (\ref{eq19}):
\beq\label{eq19a}
I_{1}=\int~\tau^2~~ds;~ I_{2}=\int~\tau^{2}~K~~ds;~
I_{3}=\int~[\tau_{s}^{2}+\tau^{2} K^{2}-~\frac{1}{4}\tau^{4}]~~ds;....
\eeq
  Here,  the total {\it curvature} $I_{0}=\int~K~ds$
  is also conserved.  This
 has no counterpart  among the conserved densities (\ref{eq18}) of
  the NLS equation.\\
{\bf (III)} Finally, setting $q=\chi$ in (\ref{eq7}) yields
 the following second analog of the DB equations:
\beq\label{eq21}
K_u=-\tau_{ss}-\frac{1}{2}(K^{2}+\tau^{2})\tau;~~ \tau_u= K_{ss}
  + \frac{1}{2}(K^{2}+\tau^{2})K.
\eeq
 Next, setting $q=\chi$ in Eq. (\ref{eq18}), we get the
   third set of infinite geometric constraints:
\beq\label{eq22}
I_{1}=\int(K^2~+\tau^{2})~ds;~I_{2}=
\int(K_{s}\tau~-K\tau_{s})~ds;~
I_{3}=\int[K_{s}^{2}+\tau_{s}^{2}-
\frac{1}{4}(K^{2}+\tau^{2})]~ds;....
\eeq
 We end this section with the remark that the total length
 of the curve, $L~=~\int~ds$ is also conserved in all three formulations,
 since the curves are non-stretching.
\section{Construction of the three curves using\\ the
  Landau-Lifshitz  equation}
\setcounter{equation}{0}
  A  general solution of the NLS (Eq. (\ref{eq7})) is of the form
 $q=\rho \exp [i\theta]$.
 Equating this  with the three complex functions  defined
 in Eqs.(\ref{eq6}), (\ref{eq11}) and (\ref{eq14}) will yield
  the   curvature and the torsion  of the three  space curves
   to be
${\bf (I)}~ \kappa_{1}=\rho,~~ \tau_{1}= {\theta}_s,~~
{\bf (II)}~ \kappa_{2}= {\theta}_s,~~\tau_{2}=\rho$ and $
{\bf (III)}~ \kappa_{3}=\rho \cos \theta,~~\tau_{3}= \rho \sin \theta$.
These are  clearly, three distinct space curves,
 each with a known curvature and torsion.
  However, {\it solving} the Frenet-Serret equations (\ref{eq1})
 explicitly for ${\bf t}$ by using these, to subsequently
  construct the position vector ${\bf r}(s,u)=\int {\bf t}~~ds$
  of  the (evolving) space curve, is
 a nontrivial task in practice.
 In the present  context, we shall  show that there is
  a certain link between the  three curve evolutions
  and the  integrable  Landau Lifshitz (LL) equation \cite{land},
 which provides us with an alternative  procedure to
 construct  the  three   moving curves.

We proceed by  first  equating
  the expressions for $\gamma_1$, $\gamma_2$ and
$\gamma_3$ given  in  Eq. (\ref{eq17})  with  those
 given in Eqs. (\ref{eq10}),
 (\ref{eq13}) and (\ref{eq16}). This yields
 the following curve evolution parameters $g, h$ and $\tau_0$
 in the three  cases:
\beq\label{eq23}
{\bf (I)}~~g_{1}=-\kappa_{1}\tau_{1};~~
h_{1}=\kappa_{1s};~~
\tau_{01}=
(\kappa_{1ss}/\kappa_{1})-\tau_{1}^{2},
\eeq
\beq\label{eq24}
{\bf (II)}~~ g_{2}=(\tau_{2ss}/\tau_{2}) -\kappa_{2}^2;~~
 h_{2}=-\tau_{2s};~~
\tau_{02}=-\kappa_{2}\tau_{2},
\eeq
\beq\label{eq25}
{\bf (III)}~~g_{3}=-\tau_{3s};~~~
h_{3}=(1/2)(\kappa_{3}^{2}+\tau_{3}^{2});~~
\tau_{03}=\kappa_{3s}.
\eeq
 Substituting  these   expressions
 for each of the  formulations  appropriately
  in Eqs. (\ref{eq2}), and using Eq. (\ref{eq1}), a short
 calculation \cite{muru} shows that
 the  completely integrable LL equation \cite{land}
\beq\label{eq26}
{\bf S}_u = {\bf S} \times {\bf S}_{ss};~~~ {\bf S}^{2}=1
\eeq
 is obtained in {\it every} case, i.e.,
 Eq. (\ref{eq26}) is satisfied by the {\it tangent} ${\bf t}_{1}$
  of the moving space curve in the first formulation, by the
   {\it binormal} ${\bf b}_{2}$
  in the second, and  the  {\it normal}
  ${\bf n}_{3}$  in the third \cite{fn}.
 Of the above, the first  may be regarded as  the
  converse  of Lakshmanan's
 mapping \cite{laks}  where, starting with the LL equation,
 and identifying ${\bf S}$ with the tangent to a moving curve,
 one   obtained the DB equations, and from them, the  NLS for
 $\psi$. The converses of the other two will yield the two analogs
 of the DB equations obtained in Section 3, and
  clearly lead to new geometries connected with the NLS.

 The LL equation (\ref{eq26})  has been shown
 to be completely integrable \cite{takh} and gauge equivalent\cite{zakh}
 to the NLS.  Its exact solutions  can be found \cite{takh,tjon}.
 We now outline \cite{muru1} how ${\bf r}_1$, ${\bf r}_2$ and
 ${\bf r}_3$, the position vectors generating the three moving curves
 underlying the NLS, can be found in terms of  an exact solution ${\bf S}$
 of Eq. (\ref{eq26}). Let $({\bf t}_i,{\bf n}_i,{\bf b}_i)$,
 $i=1,2$ and $3$, respectively,  denote the Frenet triads of the
 three moving curves  satisfying Eqs. (\ref{eq1}),  with
 the corresponding parameters $\kappa_i$ and $\tau_i$.\\
\noindent {\bf (I)}  Here,  ${\bf t}_1 = {\bf r}_{1s} $ is the
 tangent to a certain moving curve  created by a position
 vector ${\bf r}_1(s,u)$. Here, we set ${\bf t}_{1}={\bf S}$,
  a known solution of Eq. (\ref{eq26}).Thus
\beq\label{eq27}
\kappa_1=|{\bf t}_{1s}|=|{\bf S}_s|;~~\tau_1=\frac{{\bf t}_1.
({\bf t}_{1s}\times{\bf t}_{1ss})}{{\bf t}^2_{1s}}=\frac{{\bf S}.
 ({\bf S}_{s}\times{\bf S}_{ss})}{{\bf S}^2_{s}}
\eeq
Thus the underlying moving curve ${\bf r}_1(s,u)$
   is  given  by
\beq\label{eq28}
{\bf r}_1(s,u) = \int{{\bf t}_1~ ds} = \int{{\bf S}(s,u)~ds}
\eeq
  The above expression for ${\bf r}_1$  coincides with the
 surface that one would obtain using Sym's \cite{sym}
soliton-surface method.\\
\noindent {\bf (II)}  Here,  ${\bf b}_2$ is  the
  binormal of some moving curve ${\bf r}_2(s,u)$.
 We set ${\bf b}_{2}={\bf S}$.
 We have,  ${\bf t}_{2} =  {\bf r}_{2s}$.
  The curvature $\kappa_2 = \tau_1$ and torsion
 $\tau_2=\kappa_1$  (See Eq.(\ref{eq27})).
 The position vector ${\bf r}_{2}(s,u)$ generating
the  moving curve  can be shown to be \cite{muru1}
\beq\label{eq29}
{\bf r}_2(s,u) = \int{{\bf t}_2~ds} = \int{{\bf S}_{s}\times\frac{{\bf S}}
{\kappa_{1}}~~ds}
\eeq
\noindent {\bf (III)} Here,  ${\bf n}_3$
the  normal of some other moving curve ${\bf r}_{3}(s,u)$.
Thus   we set ${\bf n}_{3}={\bf S}$. Further, ${\bf t}_{3}={\bf r}_{3s}$.
  Next, from the  Frenet-Serret equations (\ref{eq1}) for this triad,
\beq\label{eq30}
(\kappa_3^2 + \tau_3^2){\bf t}_3 = \tau_3({\bf n}_3\times{\bf n}_{3s}) - \kappa_3{\bf n}_{3s}
\eeq
 A short calculation shows \cite{muru1} that
$\kappa_3 =\kappa_1 \cos \eta_1$  and
$\tau_3 =\kappa_1~~\sin \eta_1$, where
 $\eta_1=[\int \tau_1~~ds~~+~~c_{1}(u)]$.
Here, $c_{1}(u)$ is  a function of time $u$, which
 can be   determined in terms
of $\kappa_1$ and $\tau_1$, after a short calculation, by   using
 the equations for $\kappa_{3u}$ and $\tau_{3u}$ obtainable
  from  Eqs. (\ref{eq8}). Then,  substituting  the values
 for $\kappa_3$ and $\tau_3$
 into Eq. (\ref{eq30}), and setting ${\bf n}_3 = {\bf S}$,
  the position vector ${\bf r}_{3}(s,u)$
  creating the third moving space curve  can be found
 to be \cite{muru1}
\beq\label{eq31}
{\bf r}_{3}(s,u)= \int {\bf t}_3~ds=\int\frac
{[({\bf S}\times{\bf S}_s)\sin\eta_1
 - {\bf S}_s~~\cos\eta_1])}{\kappa_{1}}~ds
\eeq

\section {Moving space curves associated with the NLS soliton.}
\setcounter{equation}{0}
  As seen in the last section, the expressions for the
   three moving curves associated with the NLS
 are given in (\ref{eq28}), (\ref{eq29})
 and (\ref{eq31}) respectively, in terms of a known
 solution ${\bf S}$ of the LL equation (\ref{eq28}).
 By  defining  three orthogonal unit vectors
${\bf\hat{e}}_1=\Big\{1,0,0\Big\},~{\bf\hat{e}}_2=
\Big\{0,\cos\eta,\sin\eta\Big\},~{\bf\hat{e}}_3=\Big
\{0,-\sin\eta,\cos\eta\Big\}$,
a soliton solution  \cite{tjon} of  (\ref{eq28})
 can be written in the form
\beq\label{eq32}
{\bf S}(s,u) = (1-\mu\nu {\rm sech}^2(\nu\xi)){\bf\hat{e}}_1
+\mu\nu {\rm sech}(\nu\xi)\tanh(\nu\xi){\bf\hat{e}}_2-\mu\lambda
{\rm sech}(\nu\xi){\bf\hat{e}}_3,
\eeq
where,  $\xi = (s-2\lambda u),~ \eta = (\lambda s +
(\nu^2 - \lambda^2)u)$ and
$\mu = 2\nu/(\nu^2 + \lambda^2)$.  Here, $\nu$ and $\lambda$ denote
 arbitrary constants.
 As is clear from our discussion in Section 4, the three
 curves that result from this
 solution are associated with the soliton solution
 $q=2\nu{\rm sech}(\nu\xi)~\exp{(i\eta)}$
 of the NLS (Eq. (\ref{eq7})). They are found as follows:\\
\noindent {\bf(I)}~ Substituting Eq. (\ref{eq32}) in Eq.
(\ref{eq27}), we get
$\kappa_1 =2 \nu {\rm sech}(\nu \xi)$ and $\tau_1 =\lambda.$
Next, substituting Eq. (\ref{eq32}) in Eq. (\ref{eq28}) yields
\beq\label{eq35}
{\bf r}_1 =
\Big(s-\mu\tanh(\nu\xi),-\mu{\rm sech}(\nu\xi)\cos\eta,
 -\mu{\rm sech}(\nu\xi)\sin\eta\Big)
\eeq
This is seen to agree with the result obtained
 in \cite{levi} using Sym's \cite{sym} procedure.
In Fig. (1), we have presented a stroboscopic plot
 of the  moving curve (\ref{eq35}) at different instants
 of time. This describes the propagation of the
 well known Hasimoto "loop" soliton. In between the times at which
 we have plotted the curve, the loop changes its size and
 also rotates about the asymptotic direction which  the curve
  takes on at $s\rightarrow\pm~\infty$. In the figure, we have not plotted
 these intermediate times for the sake of clarity. The loop
 regains its form at subsequent times, as we see from the figure.\\
\noindent{\bf (II)}~ Here, we get
 $\kappa_2 = \lambda$ and $ \tau_2=2\nu {\rm sech}(\nu\xi)$.
Substituting Eq. (\ref{eq32})  in  Eq. (\ref{eq29}), the moving curve
  is  found to be
\beq\label{eq44}
{\bf r}_2=\int [\mu\lambda{\rm sech}(\nu\xi){\bf\hat{e}}_1
-\mu\lambda\tanh(\nu\xi){\bf\hat{e}}_2
+\frac{(\lambda^2-\nu^2)}{(\lambda^2+\nu^2)}{\bf\hat{e}}_{3}]~ ds
\eeq
 In this case, since the curvature is constant and the
torsion vanishes as  $s\rightarrow\pm\infty$, for {\it all}
 finite times, the curve is bounded by two planar circles at both
 ends. The maximum non-planarity of the curve occurs
 at $s=2\lambda u$.
 A stroboscopic plot of Eq. (\ref{eq44}) given
 in Fig. (2). This curve is clearly seen to rotate
   and propagate as time progresses. \\
\noindent{\bf (III)}~ Here, $\kappa_3 =
 2\nu{\rm sech}(\nu\xi)\cos\eta$ and $\tau_3 =
 2\nu{\rm sech}(\nu\xi)\sin\eta$.
  When Eq. (\ref{eq32}) is substituted
 in Eq. (\ref{eq31}), with $\eta_1=\eta$, a long but straightforward
 calculation yields
\bea\label{eq46}
\lefteqn{{\bf r}_3 =\int[\mu{\rm sech}(\nu\xi)(\lambda\sin\eta-\nu\tanh(\nu\xi)\cos\eta)
{\bf\hat{e}}_1 {} }\nonumber\\
& & {}-(\mu\lambda\tanh(\nu\xi)\sin\eta+(1-\mu\nu\tanh^2(\nu\xi))\cos\eta){\bf\hat{e}}_2 {} \nonumber\\
& & {}+(\frac{(\lambda^2-\nu^2)}{(\lambda^2+\nu^2)}\sin\eta -\mu\lambda\tanh(\nu\xi)\cos\eta){\bf\hat{e}}_{3}]~ ds
\eea
 A stroboscopic plot of the moving curve (\ref{eq46})
 at different instants of time is given
  in Fig. (3). This depicts  the propagation of a "loop" soliton
 distinct from the Hasimoto soliton in the sense that
 it also oscillates with time,  as
 is clear from the figure. Furthermore, at intermediate
 times, the loop rotates about the asymptotic direction,
 decreasing its loop size, and almost straightens out
 periodically, after which it starts  looping  again.
 These intermediate plots have not been presented in
 the figure.

\section{Three curve velocities associated with
 NLS evolution}
\setcounter{equation}{0}
 Returning to our general results of  Section 4,
  the three position vectors ${\bf r}_{i}$, $i=1,2$  and $3$,
  which are associated with the NLS evolution, can be found from Eqs.
 (\ref{eq28}), (\ref{eq29}) and (\ref{eq31}), respectively,
  using an exact solution${\bf S}$ of the LL equation (\ref{eq26}).
The corresponding
 curve velocities ${\bf v}_{i}(s,u)= {\bf r}_{iu}$  at each
 point $s$, can  therefore be
 found from these  equations  by  direct
 differentiation with respect to time, $u$.
 However, to compare and contrast the  intrinsic geometries
  of the three space curves, it is
  instructive to express these
 velocities in terms  of  the vectors of the
 corresponding  Frenet triads, as well as the curve
 parameters, as follows.
  Since the curves are non-stretching, we have ${\bf v}_{is}= {\bf r}_{ius}=
 {\bf r}_{isu}={\bf t}_{iu}$.
 On the other hand, from the first equation of Eqs. (\ref{eq2}),
 we have, ${\bf t}_{iu}=g_{i}{\bf n}_{i}+ h_{i}{\bf b}_{i}$.
  Here, the quantities $g_{i}$ and $h_{i}$
 for the three curves of the NLS are given in
 Eqs. (\ref{eq23}), (\ref{eq24}) and (\ref{eq25}), respectively.
 Using these in ${\bf v}_{i}(s,u)= ~\int~{\bf t}_{iu}~~ds$, we get
\beq\label{eq36}
{\bf (I)}~~{\bf v}_1=~\int (-\kappa_{1}
 \tau_{1}{\bf n}_{1} + \kappa_{1s} {\bf b}_{1})~ds=
~\kappa_{1}~{\bf b}_{1}.
\eeq
\beq\label{eq37}
{\bf (II)}~~{\bf v}_{2}=~\int~[(\tau_{2ss}/\tau_{2}-
~\kappa_{2}^{2}){\bf n}_2-\kappa_{2s}{\bf b}_{2}]~ds.
\eeq
\beq\label{eq38}
{\bf (III)}~~{\bf v}_{3}=~\int~[-\tau_{3s}{\bf n}_{3}
 +~\frac{1}{2}(\kappa_{3}^{2} +\tau_{3}^{2}){\bf b}_3 ]~ds.
\eeq
  As expected,  ${\bf v}_{1}$ coincides
 with the vortex filament velocity Eq. (\ref{eq3}) derived by Da Rios
   in fluid mechanics.
 It is  a {\it local} expression
 in the curve variables: The velocity ${\bf v}_{1}$ at a point
 $s$ depends on the curvature and binormal at that point only.
 In contrast, ${\bf v}_{2}$ and ${\bf v}_3$ are  seen to be
{\it nonlocal} in the curve variables: If we  partially integrate
 the right hand sides of  Eqs. (\ref{eq37}) and (\ref{eq38}), and
 use the Frenet-Serret
  equations (Eqs. (\ref{eq1})) repeatedly
 in the resulting expressions,
 both these velocities take on the form
 ${\bf v}_{i} = A_{i}~{\bf t}_{i}
 +~ B_{i}~{\bf n}_{i}+~C_{i}~{\bf b}_{i}$, $i=2,3$,
 where the components $A_{i},B_{i}$ and $C_{i}$
   can be written in terms of {\it integrals} of
  certain functions involving
 the curvature,  the torsion , their
 higher derivatives, and their various products.
 It is indeed interesting that in spite  of such a
 complicated behaviour of these velocities,
 their  corresponding  curve evolutions are also endowed with an
infinite number of constants of motion.
This essentially  stems from their connection with the NLS.
 More specifically, this  is because, as we have shown in Section 3,
   the binormal ${\bf b}_{2}$ of  the second curve
and the normal ${\bf n}_3$  of  the third, satisfy the integrable
 LL equation.

 We conclude with the following remarks:
In a fluid, as is well known,
 the induced  velocity ${\bf v}$
 at  a point is determined  as a volume integral
  involving  the vorticity
 ${\bf \Omega}$, by using the Biot-Savart formula.
 First, it is to be  noted that in this formula, if one  expresses
 ${\bf \Omega}$ in terms of the
 vectors of the Frenet triad of the filament,
   then, in any realistic model of a fluid,
 ${\bf v}$ {\it is}  nonlocal in the curve variables, and becomes local only under certain
 approximations.  Secondly, it is worth noting that in
 an interesting
 experiment with a fluid in a rotating tank, Hopfinger
 and Browand
\cite{hopf} had observed  {\it compact} distortions which twist and
 propagate  along a vortex core  like a  soliton.
  These two features  suggest that
 our new results  which link the velocities
 ${\bf v}_{2}$ and ${\bf v}_{3}$
 with the NLS and its soliton, may be of
 some relevance in actual fluids.  In view of this, appropriate
  theoretical modeling of the vorticity, as well as possible
 experimental  studies of the detailed geometric structure
 of moving vortex filaments, to look for any similarities
 with Figures (1) to (3) (which depict
 propagation of compact distortions) would indeed be of interest.

\newpage
\begin{figure}
\resizebox{0.5\textwidth}{!}{%
\includegraphics{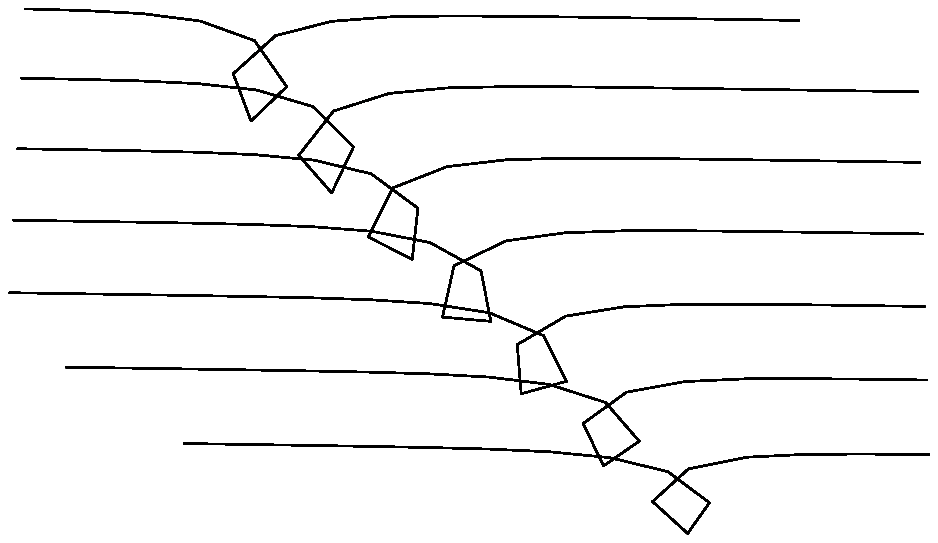}
}
\caption { A stroboscopic plot of  the evolving
 space curve ${\bf r}_{1}(s,u)$ (Eq. (\ref{eq35}))
 for $\nu=1$  and $\lambda=0.1$.}
 \end{figure}
\begin{figure}
\resizebox{0.5\textwidth}{!}{%
\includegraphics{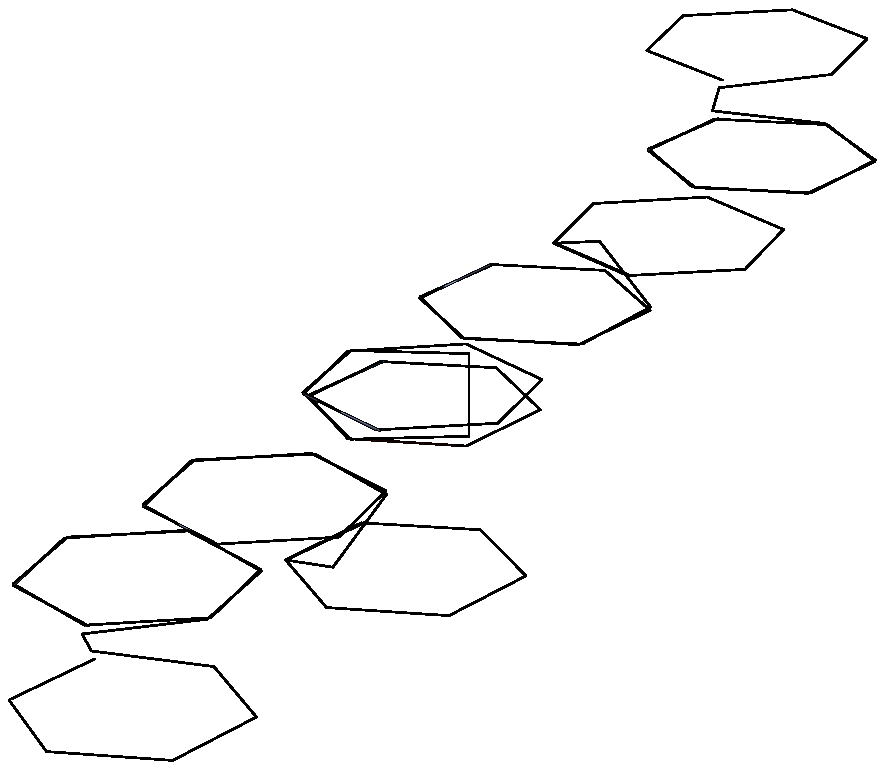}
}
\caption {A stroboscopic plot of the evolving
 space curve ${\bf r}_{2}(s,u)$ (Eq. (\ref{eq44}))
 for $\nu=0.3$ and $\lambda=0.1$.}
\end{figure}
\begin{figure}
\resizebox{0.5\textwidth}{!}{%
\includegraphics{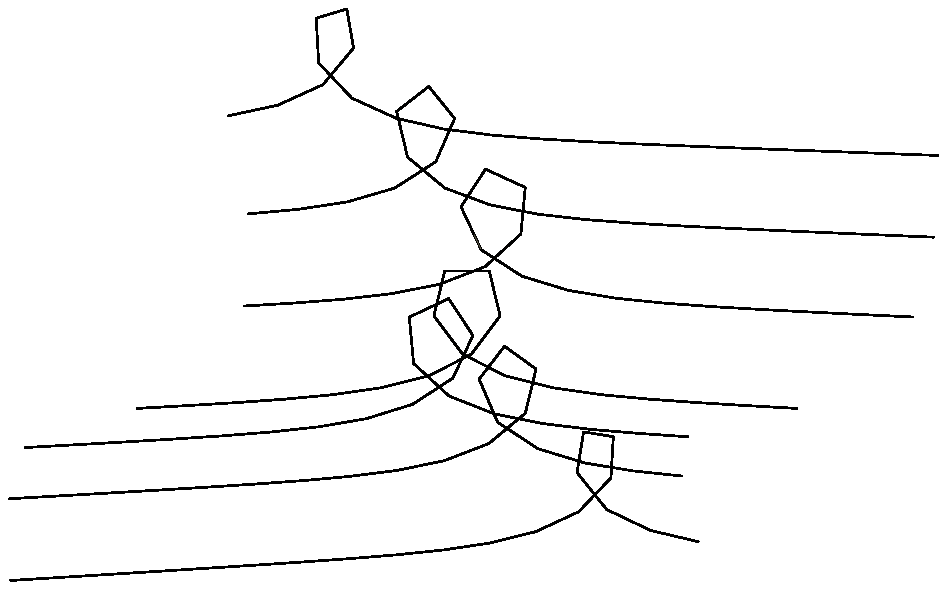}
}
\caption {A stroboscopic plot of  the evolving
 space curve ${\bf r}_{3}(s,u)$ (Eq. (\ref{eq46}))
 for $\nu=1$ and $\lambda=0.3$.}
\end{figure}

\begin{thebibliography}{99}
\bibitem{hami} Here, $I_{3}$ is the Hamiltonian which gives
 rise to (\ref{eq7}), on using the Poisson brackets $\{F,G\}=i\int~[(\delta F/\delta q)
(\delta G/\delta q^*)~-~(\delta G/\delta q)(\delta F/\delta q^*)]~ds$.
\end{thebibliography}
\end{document}